\documentclass[twocolumn,floatfix]{aastex61}
\usepackage{url}
\usepackage{natbib}
\usepackage{booktabs}
\usepackage{epstopdf}
\usepackage{graphicx}
\journalinfo{\textsc{Accepted in The Astrophysical Journal}}
\usepackage{array,etoolbox}

\preto\tabular{\setcounter{magicrownumbers}{0}}
\newcounter{magicrownumbers}
\newcommand\rownumber{\stepcounter{magicrownumbers}\arabic{magicrownumbers}}

\shorttitle{Magnetic imprints of eruptive and non-eruptive solar flares}
\shortauthors{Vasantharaju et al}
\begin{document}
	\title{Magnetic imprints of eruptive and non-eruptive solar flares as observed by Solar Dynamics Observatory}
	\author{N.~Vasantharaju}
	\email{vrajuap@gmail.com}
	\affil{Indian Institute of Astrophysics, Koramangala, Bengaluru-560 034, India}
	\affil{Dipartimento di Fisica e Astronomia ``Ettore Majorana'' - Sezione Astrofisica, Universita degli Studi di Catania, Via S. Sofia 78, I-95123 Catania, Italy}
	\author{P.~Vemareddy}
	\affil{Indian Institute of Astrophysics, Koramangala, Bengaluru-560 034, India}
	\author{B.~Ravindra}
	\affil{Indian Institute of Astrophysics, Koramangala, Bengaluru-560 034, India}
	\author{V.~H.~Doddamani}
	\affil{Department of Physics, Bangalore University, Bengaluru-560 056, India}
	%%%%%%%%%%%%%%%%%%%%%%%%%%%%%%%%%%%%%%%%%%%%%%%%%%%%%
	%% Abstract %
	%%%%%%%%%%%%%%%%%%%%%%%%%%%%%%%%%%%%%%%%%%%%%%%%%%%%%
\begin{abstract}
The abrupt and permanent changes of photospheric magnetic field in the localized regions of active regions during solar flares called magnetic imprints (MIs), have been observed for the past nearly three decades. The well known ``coronal implosion'' model is assumed to explain such flare associated changes but the complete physical understanding is still missing and debatable. In this study, we made a systematic analysis of flare-related changes of photospheric magnetic field during 21 flares (14 eruptive and 7 non-eruptive) using the high-cadence (\texttt{135s}) vector-magnetogram data obtained from Helioseismic and Magnetic Imager. The MI regions for eruptive flares are found to be strongly localised, whereas the majority of non-eruptive events ($>70~\%$) have scattered imprint regions. To quantify the strength of the MIs, we derived the integrated change of horizontal field and total change of Lorentz force over an area. These quantities correlate well with the flare strength, irrespective of whether flares being eruptive or not, short or long duration. Further, the free-energy (FE), determined from virial-theorem estimates, exhibits statistically significant downward trend which starts around the flare time is observed in majority of flares. The change of FE during flares do not depend on eruptivity but have a strong positive correlation ($\approx 0.8$) with the Lorentz force change, indicating that the part of FE released would penetrate into the photosphere.  While these results strongly favor the idea of significant feedback from corona on the photospheric magnetic field, the characteristics of MIs are quite indistinguishable for flares being eruptive or not.
\end{abstract}
%However, the physical significance of many observations related to magnetic imprints are still elusive. of photospheric magnetic field We observed the change of Lorentz force corresponding to the rapid and permanent changes of the horizontal magnetic field during all flares in our sample.
	
	%%%%%%%%%%%%%%%%%%%%%%%%%%%%%%%%%%%%%%%%%%%%%%%%%%%%%%%%%
	%% 1. Introduction %
	%%%%%%%%%%%%%%%%%%%%%%%%%%%%%%%%%%%%%%%%%%%%%%%%%%%%%%%%%
\section{Introduction}
	\label{Intro}
%	\linenumbers
	
%\pv{From now on, whenever you submit a paper, the text is cross checked with a software. The check is on plagiarism, if same sentences are found as is from previous articles, then the article will be reverted to the authors. You please modify if taken as is. Generally I prepare text with my own framing of sentences, depending little on previous paper text. }

%\pv{somewhere in the appropriate context, cite Maurya et al 2012 ApJ 747, 134 and S. Gosain 2012 ApJ 749 85. Both can be fitted here. }

%\pv{Bibliography should reduce too many authors. Citation with More than 3 authors, can be changed with et al. This has to be done in .bib file. This I had told you some time earlier.}

The magnetic field is believed to be the viable source of energy for the released thermal and radiant energy in solar flares and the observed kinetic energy in CMEs \citep{Forbes2000,Hudson2007}. It is widely accepted that solar eruptions like flares and CMEs are the result of coronal magnetic restructuring, generally involving magnetic reconnection, during which the magnetic free energy gets converted into heat, electromagnetic radiation and kinetic energy \citep{Priest2002}. More than three decades of observational studies (e.g. \citealt{Patterson1981,Wang1992,Wang2012b}) confirm that the solar photosphere responds to the sudden coronal magnetic restructuring that occurs during most major flares. The flare associated changes in photospheric magnetic field  can be 'transients' (e.g. \citealt{Maurya2012}) and 'permanent' as well. In this article, we are discussing abrupt, permanent and widely distributed patterns of magnetic field change associated with flares called magnetic imprints. Here, the magnetic field change is considered ``permanent'' if its effect lasts for at least couple of hours after the flare.
	
 As far as the history of studies on magnetic imprints are  concerned, \citet{Wang1992} provided the first clear illustration of it using ground based vector magnetograms. They reported the step-wise enhancement of magnetic shear co-temporal with the initiation of X3 class flare. Later \citet{Kosovichev2001} with the help of the line-of-sight (LoS) magnetograms obtained from Michelson Doppler Imager (MDI; \citealp{Scherrer1995}) on board the Solar and Heliospheric Observatory (SOHO; \citealp{Domingo1995}) satellite, observed the abrupt and permanent magnetic flux decrease near the polarity inversion lines during a X-class flare occurred near the disk center. \citet{Zharkova2005} observed LoS magnetic flux increase during a X-class flare occurred near the limb. Generally, the LoS magnetic flux increases in the active regions (ARs) located near the limb and decreases in ARs near to the disk center indicating the more horizontal magnetic field configuration \citep{Wang2002}. By availing LoS magnetograms obtained from Global Oscillations Network Group (GONG) ground-based network, \citet{Sudol2005} found the abrupt, significant and permanent changes in longitudinal magnetic field strength during 15 X-class flares. By analysing vector magnetograms obtained from the ground based vector magnetographs, \citet{Wang2010} studied 11 X-class flares and found an increase of transverse field at the PIL of each event source AR.
 
 With the launch of Solar Dynamics observatory, the continuous, high spatial and temporal resolution full disk vector magnetograms are being obtained from the Helioseismic and Magnetic Imager (HMI; \citealp{Schou2012}) instrument on board  Solar Dynamics Observatory (SDO; \citealp{Pesnell2012}). Using 12 minute vector magnetograms, rapid and irreversible magnetic imprints  on the photospheric  magnetic field i.e., leading to more horizontal magnetic field structure, were observed during several major eruptions \citep{Petrie2012,Petrie2016,Zekun2019}. \citet{Wang2012b} observed magnetic imprints on the photospheric magnetic field during 18 different GOES class flares (includes C-class events) and found that the vertical Lorentz force changes during flares were generally directed downward near PILs and the absolute value of these forces are well correlated with the flare strength. \citet{Sun2017} introduced the high cadence (90 s and 135 s) vector magnetograms from the HMI to characterize the rapid evolution of horizontal magnetic field.
 %, while the routine version of 720 s cadence vector magnetograms from HMI are not that effective to record such rapid variations.
 
 Nevertheless, the physical understanding of flare associated field changes in photosphere is unclear yet. \citet{Hudson2000} proposed magnetic implosion conjecture to explain such flare associated changes in photosphere. The conjecture claims that magnetic implosion must occur simultaneously with coronal magnetic restructuring during flares/CMEs led to the abrupt contraction of loop structure makes the photospheric magnetic field to become ``more horizontal''~\citep{Hudson2008}. Further developing on this model, \citet{Fisher2012} claimed that the perturbation of Lorentz force balance results in upward driving of CME mass to a greater speed and simultaneously, the same amount of impulse must act downwards on the photosphere due to momentum conservation. This downward impulse also known as 'magnetic jerk' is believed to drive the helioseismic waves into the solar interior (``Sunquakes'' e.g., \citealt{Kosovichev1998}). Another possible mechanism responsible for the generation of sun-quake is by the strong hydrodynamic shocks travelling downward to photospheric levels and solar interior at supersonic speed, due to the injection of energetic particle beams into the chromosphere in flaring atmosphere \citep{Fisher1985,Zharkova2015}.

%\pv{First give the definition of magnetic imprint before it being used.}

%\pv{Don't use citealp, we need to cite chronologically. That is reasonable}.

Further, the observations of contractions of coronal loop structures during major flares in coronal EUV images \citep{Gosain2012,Sun2012,Liu2012,Simoes2013,Wang2018} provides the clear evidence of coronal implosion as predicted by \citet{Hudson2000}. Though we have such a vast accumulation of observational evidences favouring coronal implosion model, the origin of magnetic imprints is still not clear i.e., How the surface magnetic field becomes more horizontal? Also, the recent observational studies \citep{Sun2017,Kleint2017} raised many questions on the validity of the coronal implosion model such as the photospheric magnetic field would not respond in a similar manner as that of contracting coronal field during flares, and ubiquitous observations of magnetic imprints all over the AR leads to the neutralisation of oppositely directed vertical Lorentz forces and the resultant upward impulse is insufficient to drive the CME outwards.

In the present study, we analysed 21 flare events in a systematic way by availing the \texttt{135\,s} cadence vector magnetograms and compared them quantitatively to check the validity of the prediction  made by \citet{Hudson2008} and \citet{Fisher2012}. Meanwhile, we also attempted to address a few questions pertaining to magnetic imprints in order to understand the back reaction during flares effectively. Thus, the motivation of this study is threefold: 1. To check the validity of the conjecture \citep{Hudson2008,Fisher2012} for large number of different GOES class flare events. 2. Whether the eruptive flares yield larger magnetic field changes (strength of magnetic imprints) on the photosphere compared to non-eruptive counterparts? 3. Does the amount of decrease in magnetic free energy during flares have any correspondence with the back reaction on solar surface? Details of the observational data and methodology are given in Section ~\ref{ObsData}. The analysis and results are described in Sections ~\ref{Res}. Summary and discussions are given in Section ~\ref{summ}.

\begin{figure*}[!ht]
	\centering
    \includegraphics[width=15cm,height=14.7cm,clip=]{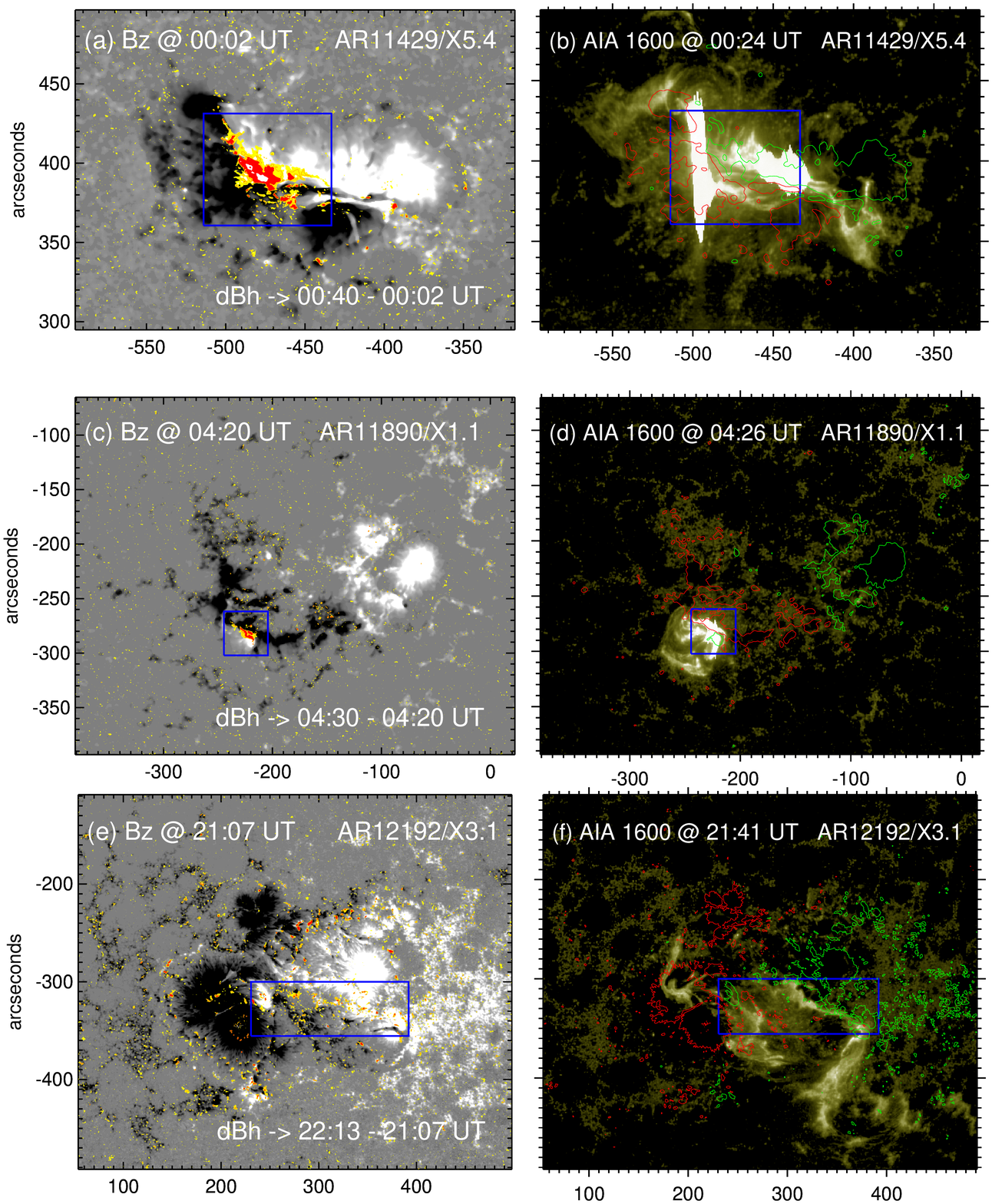}
	\includegraphics[width=14.68cm,height=4.5cm,clip=]{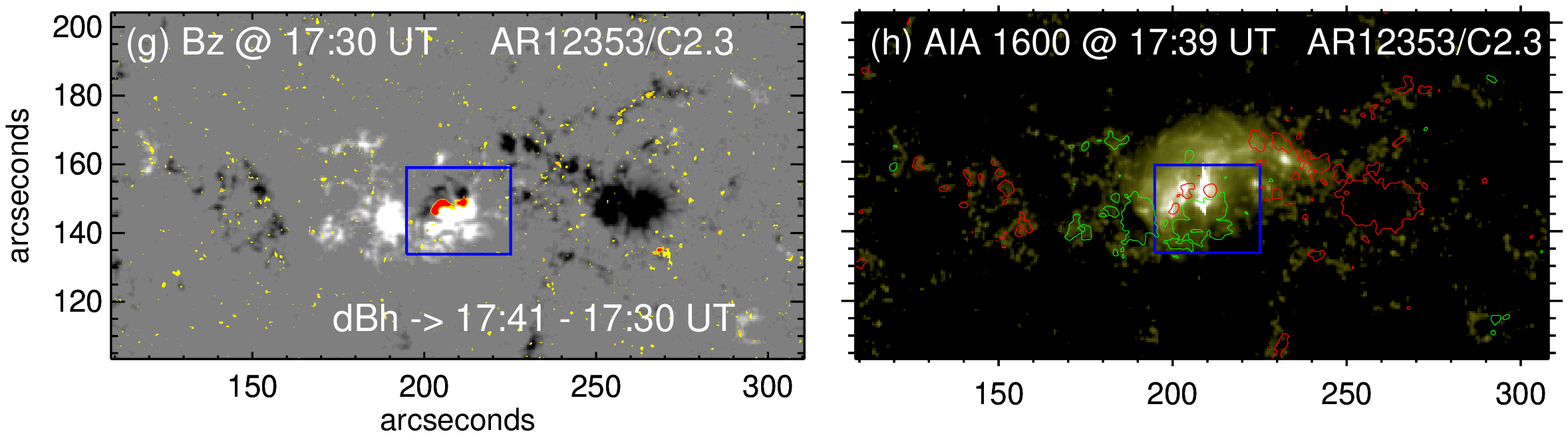}
	\caption{Illustration of four sample events  to identify the regions (ROIs) of the magnetic imprints. Top two events (a-d) are eruptive and bottom two (e-h) are non-eruptive events. {\bf (a-b):} Eruptive flare event of X5.4 from AR 11429. The $B_z$ map is plotted in panel (a), over which the enhancements of $B_h$ between the flare initial time (00:02UT) and end time (00:40 UT) are plotted. The yellow, orange and red filled contours represent the regions with $B_h$ enhancements of 200 G, 300 G and 400 G respectively. Panel (b) is the AIA 1600~\AA~image observed at flare peak time (00:24 UT), overlaid are the $B_z$ contours  at $\pm 600~G$. The colored contour region within blue box in (a) marks the ROI, where the flare kernels in panel (b) are spatially coherent with the $B_h$ enhancements. The region enclosed blue box, which exclude non-flare associated changes, is used for further computations.{\bf (c-d):} Short duration eruptive event of X1.1 from AR 11890. Similarly, {\bf (e-f):} for non-eruptive event of  X3.1 from AR 12192, where magnetic imprints are scattered. {\bf (g-h):} non-eruptive event of  C2.3 from AR 12353
	%\pv{naming the panels a) but it is A). Generally we use small letters to refer in the text} 
	}
	\label{fig1}
\end{figure*}

%%%%%%%%%%%%%%%%%%%%%%%%	
	
	%\begin{longrotatetable}
\begin{deluxetable*}{lllllllcccccc}[!htb]
\centering	
		\tablecaption{List of 21 flare events from 17 ARs and their associated magnetic properties
		\label{Tab1}}
%		\tablewidth{700pt}
		\tabletypesize{\scriptsize}
		\tablehead{\colhead{S.No} &
			\colhead{NOAA} & \colhead{Date} & 
			\colhead{Loc} & \colhead{GOES} & 
			\colhead{$T_{ini}$} & \colhead{$T_{end}$} & 
			\colhead{$ Int.~ \delta B_{h}$ } & \colhead{$Tot.~ \delta F_{z}$ } & 
			\colhead{$\Delta FE $ }  & \colhead{$T_{cme}$} & \colhead{$V_{cme}$} & \colhead{$M_{cme}$} \\ 
			\colhead{} & \colhead{AR} & \colhead{} & \colhead{} & \colhead{class} & \colhead{(UT)} & \colhead{(UT)} &
			\colhead{$(10^{20} Mx)$} & \colhead{$(10^{22}dyne)$ } & \colhead{$(10^{31}ergs)$}  & \colhead{(UT)} & \colhead{(km s$^{-1}$)} & \colhead{$(10^{15} g)$} } 
\startdata
\rownumber & 11158     &     20110215  &       S21W14  &     X2.2 &       01:44  &     02:06  &     $6.1\pm0.8$ &    $4.7\pm1.1$   &    $5.7\pm4.5$  &   02:24 &  669 & 4.3 \\
\rownumber & 11283  &  20110906  &  N14W04  &  X2.1  &  22:12  & 22:34   &   $6.7\pm0.7$     &    $4.9\pm1.2$  &     $5.1\pm4.3$    &      23:05   &    575 & 15.0   \\
\rownumber & 11283  & 20110907   & N14W18  &  X1.8  &  22:32  & 22:44   &   $6.6\pm1.0$    &    $3.1\pm0.9$   &    $4.0\pm3.6$      & 23:05    &   792 &  1.1\\
\rownumber & 11429\tablenotemark{b} &  20120309 &   N15W03 &   M6.3  &  03:22  & 04:18   &   $4.0\pm1.0$    &    $2.2\pm1.0$   &    --  &   04:26 &   950  & 5.6\\
\rownumber & 11890  & 20131110  &  S11W15  &  X1.1  &  05:08  & 05:24   &   $5.0\pm1.0$     &    $3.1\pm0.9$   &  $2.0\pm2.3$  &   05:36   &      682 & 2.3      \\
\rownumber & 12017  & 20140329  &  N10W20 &   X1.0  &  17:35  & 18:01   &   $4.9\pm1.2$     &     $1.9\pm0.9$      & $5.0\pm4.1$   &    18:12   &  528 & 5.0 \\
\rownumber & 12158  & 20140910  &  N15E14  &  X1.6  &  17:21  & 18:20   &   $11.9\pm0.8$    &    $6.1\pm1.7$    &    $5.8\pm4.2$   &       18:00   &    1267 & 21.0  \\ 
\rownumber & 12242  & 20141220   & S18W29  &  X1.8  &  00:11  & 00:55  &    $17.3\pm1.0$    &   $9.7\pm0.7$     &     $20.2\pm8.9$ &      01:25   &  830 & 22.0  \\
\rownumber & 11719 &  20130411  &  N09E12 &   M6.5 &   06:55 &  07:29 &     $6.1\pm1.2$ &   $0.9\pm0.6$  & $3.5\pm2.9$ &   07:24 & 861 & 22.0   \\
\rownumber & 12371 &  20150622 &   N12W08 &    M6.5 &   17:39 &   18:51 &      $18.4\pm1.2$ &    $9.8\pm1.6$   &  $17.2\pm10.1$ &  18:36 &  1209 & 4.4 \\
\rownumber & 12443 &  20151104  &  N08W02 &   M3.7  &  13:31 &   14:13 &      $6.4\pm2.0$  &   $1.0\pm0.8$ & $6.5\pm5.3$ & 14:48 & 578 & 6.6  \\
\rownumber & 11429 &  20120307  &  N17E29  &  X5.4  &  00:02  &  00:40 &     $20.6\pm1.4$   &  $12.4\pm2.6$  &  $39.9\pm17.5$  & 00:24 & 2684 &  14.0  \\
\rownumber & 11890 &  20131108  &  S14E15  &  X1.1  &  04:20  & 04:30  &     $3.5\pm0.9$  &   $2.7\pm0.8$  & $9.7\pm5.8$   & 03:24  & 497 & 8.8 \\
\rownumber & 11882 &  20131028  &  S08E34  &  M4.4 &  15:07 &  15:21  &    $2.5\pm1.0$   &    $1.4\pm0.6$   & $2.2\pm2.6$ & 15:36 &  812 & 9.3   \\
\rownumber & 11166\tablenotemark{a} &  20110309  &  N08W09  &  X1.5  &  23:13  &  23:39   &   $4.5\pm1.0$   &   $2.9\pm0.6$  &  $5.0\pm4.1$  &  -- &  -- & --  \\
\rownumber & 12192\tablenotemark{a}  & 20141024  &   S16W21  &  X3.1  &  21:07  & 22:13   &   $18.3\pm3.1$     &    $9.9\pm2.4$    &   $12.3\pm18.8$ &  --  & -- &  --   \\
\rownumber & 12297\tablenotemark{a}  &  20150311  &  S17E22  &  X2.2  &  16:11  & 16:29  &    $5.3\pm1.2$    &    $5.1\pm0.8$   &    $7.8\pm5.2$  &   --  &   -- &  --   \\
\rownumber & 11261\tablenotemark{a} &	20110730  &  N16E33  &  M9.3  &  02:04  & 02:21  &    $2.6\pm0.8$  &    $2.1\pm0.7$   &   $2.1\pm1.0$  & -- & -- & -- \\
\rownumber & 11967\tablenotemark{a} &  20140202  &  S11E05  &  C9.7 &   11:45  & 11:55  &    $7.4\pm1.3$   &   $1.4\pm0.7$   & $1.7\pm6.8$   &  --   &  --  & --  \\
\rownumber & 11158\tablenotemark{a}  & 20110215  &  S20W08  &  C4.8  &  04:27 &  04:37 &     $0.8\pm0.7$  &   $0.3\pm0.6$  & $0.6\pm2.2$  & --  & -- & --    \\
\rownumber & 12353\tablenotemark{a}  & 20150523 &    N07W12 &   C2.3  &  17:30  &  17:41  &     $1.1\pm1.3$  &   $0.6\pm 0.7$ &   $0.2\pm 0.5$  &  --  &   -- &  -- \\
\enddata
\tablenotetext{a}{Non-eruptive flare events}
\tablenotetext{b}{Flare event didn't exhibit decrease in free energy.}
\tablecomments{ $T_{ini}$ and $T_{end}$ : The start and end times of the flare respectively, are obtained from GOES catalog. $ Int.~ \delta B_{h}$ : Integrated change of $B_{h}$ during the flare. $Tot.~ \delta F_{z}$ : Total change of Lorentz force during the flare. $\Delta FE $ : Decrease in magnetic free energy during the flare. $T_{cme}$ : Time of CME structure first appeared on LASCO C2 coronagraph. $V_{cme}$ and $M_{cme}$: Linear speed and mass of CMEs are obtained from the LASCO CME catalog respectively.}
\end{deluxetable*}
%\end{longrotatetable}

\section{Observational data, processing and methods}
\label{ObsData}
%\pv{Define HMI, SDO with citation in the first usage of introduction}
To study the rapid magnetic evolution during solar eruptions, we have used the 135 s vector magnetogram data (\texttt{hmi.B\_135s} series) obtained from HMI on board SDO. This data set has the same format and pixel size (0\arcsec.5) as that of the routine 720 s version of data set and is processed with identical pipeline procedures with a few exceptions, were discussed in \citet{Sun2017}. They did a detailed comparison indicating that the high-cadence data agrees well with the routine version in strong-field regions ($B > 300 G$). However, a caveat should be considered while performing analysis on transverse magnetic field in weaker-field regions due to higher noise. This high cadence data series are in native Helioprojective-Cartesian coordinates (field strength, inclination and azimuth) and these field-vectors have to be re-projected into a Heliocentric-spherical coordinates ($r,\theta, \phi$). For this, we have adapted the HMI pipeline procedures along with several modules under solar soft HMI branch to resolve the azimuth ambiguity (\texttt{hmi\_disambig.pro}) and re-mapping of full-disk data into Cylindrical Equal Area (CEA) coordinates (\texttt{bvec2cea.pro}). Finally, the obtained field components ($B_{p},B_{t},B_{r}$) in  Heliocentric spherical coordinate were approximated to Heliographic components ($B_{x},-B_{y},B_{z}$) ~\citep{Sun2013}. %To avoid the uncertainties in the measurements, we have considered the pixels greater than 50 G in $B_{z}$ distribution and 150 G in transverse magnetic field maps.
	
As the high-cadence HMI data has been processed for selected periods of flaring activity between August 2010 and November 2015, we have chosen a sample of 21 flare events from 17 ARs, in which 14 are CME associated events (eruptive) and 7 are not CME associated ones (non-eruptive/confined). We included a few C-class flare events in our sample because many times weaker flares do not show decrease in free energy owing to the uncertainties in vector magnetic field measurements \citep{Vasantharaju2019}. Hence, we accumulated more of strong flare events in our sample and all these events are located within $40^\circ$ from the central meridian of the solar disk to minimize the projection effects. The source ARs of sample flare events along with magnetic imprint properties, associated CME details are tabulated in Table~\ref{Tab1}.
	
Atmospheric Imaging Assembly (AIA; \citealp{Lemen2012}), an another instrument on board SDO provides the full disk ultra-violet (UV; 1600~\AA~ and 1700~\AA) images at the cadence of 24 s with a pixel size of $0\arcsec.6$. We have used the AIA 1600~\AA~ images to trace the morphology of flare ribbons and in turn helps in identifying the Region-Of-Interest (ROI) on the vector magnetograms. The methodology we followed in the identification of ROI, computation of Lorentz force  and Magnetic free-energy will be discussed in the following subsections.
	
\subsection{Lorentz force}
	\label{lf}
As exemplary cases, we considered four events (two eruptive and two non-eruptive) to demonstrate the procedure for computation of Lorentz force change during flares and to distinguish the morphological features of magnetic imprints (MI) between eruptive and non-eruptive events. The X5.4 and X1.1 flares are associated with CME eruptions (Figure~\ref{fig1}(a-d)), whereas X3.1 and C2.3 flares are not associated with CMEs (Figure~\ref{fig1}(e-h)). 
%The snapshots of AR 11429 and AR 11890  in AIA 1600~/AA waveband during X5.4 flare peaked at 00:24 UT on 7 March 2012 and X1.1 flare peaked at 4:26 UT on 8 November 2013. X2.2 flare peaked at 01:56 UT on 15 February 2011 occurred from AR 11158 located in the southeast of solar disk and X1.5 flare, a non-eruptive event having the peak time at 23:23 UT on 9 March 2011 occurred from AR 11166 located in the  northwest of solar disk. 

The procedure to identify the ROI enclosing the flare related changes in the vector magnetograms is similar to \citet{Wang2012b}. The rapid and irreversible increase of horizontal magnetic field, $B_h = \sqrt{B_x^2 + B_y^2} $, in the flaring region can be located approximately with the help of corresponding AIA 1600~\AA/1700~\AA~images. But to specify the exact region of magnetic imprints, we constructed the difference image of horizontal magnetic field ($\delta B_h$) observed nearest to flare initial and end times. We then over-plotted the contour of level 150 G on the $\delta B_h$. It's worth to note here that \citet{Sun2017} claims that the error in transverse magnetic field for 135 s cadence data is 50 G higher than that of the routine version of 720 s cadence vector magnetograms. Hence the error ($\sigma$) in $B_h$ for 135 s cadence data is 150 G and the propagation of error in the difference maps of horizontal magnetic field would be $\sigma \delta B_h = \sqrt{\sigma B_{h,i}^2 + \sigma B_{h,f}^2}  \approx $ 200 G. The left panels of Figure~\ref{fig1} are the $B_z$ maps at the onset of flares and the contours of $B_h$ enhancements at 200 G (yellow), 300 G (orange) and 400 G (red) are over-plotted on the respective $B_z$ maps. In the right panels of Figure~\ref{fig1}, we displayed the corresponding snapshots of AIA 1600~\AA~observed at their flare peak times. By careful manual inspection of contour regions of $\delta B_h$ and the corresponding AIA images, we then choose the ROI enclosing the contour region with level greater than 150 G. The blue boxes represent ROI enclosing the flare-related field change. In this way, the non-flare related enhancement of horizontal field is well excluded. Note that the blue boxes enclosing the flare kernels in AIA 1600~\AA~images are spatially well correlated with $B_h$ enhancements in the left panels of Figure\ref{fig1}. It is noted that the distinguishing morphological feature of MI regions produced by eruptive events from that of non-eruptive ones is that the MI regions of eruptive flares are more strongly localised than MI regions of non-eruptive events. Based on the qualitative assessment of MI regions morphology, it is found that more than $70~\%$ of non-eruptive events in our sample produces scattered regions of MI. The MI region in Figure~\ref{fig1}(e) produced by non-eruptive flare X3.1 is scattered whereas the MI regions in Figure~\ref{fig1}(a) and ~\ref{fig1}(c) produced by eruptive events of X5.4 and X1.1 are more localised. A few non-eruptive events ($<30~\%$) also produce the compact/localised MI regions (Figure~\ref{fig1}(g)). The summation of $\delta B_h$ over the whole ROI gives the value of the integrated change of $B_h$ and the values computed for sample events are tabulated in column 7 of Table~\ref{Tab1}. 

%In figure(\ref{fig1}a \&~\ref{fig1}c), we over-plotted the contours of $B_h$ enhancements at 200 G (yellow), 300 G (orange) and 400 G (red) on the $B_z$ map, respectively.In figure(\ref{fig1}b \&~\ref{fig1}d), we displayed the corresponding AIA images observed at their flare peak times.
	
\begin{figure*}[!ht] 
    \centering
   \includegraphics[width=.98\textwidth,clip=]{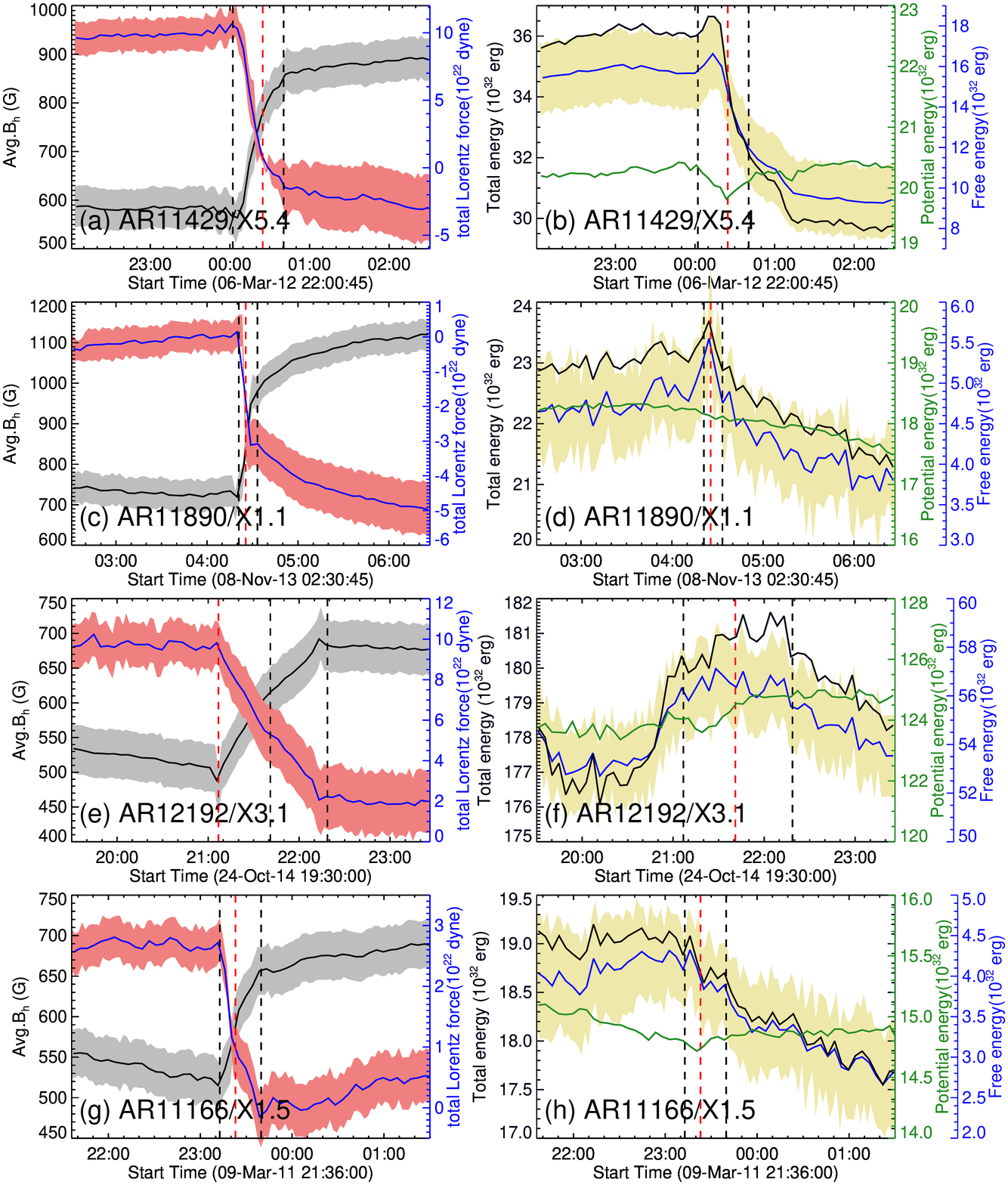}
	\caption{Temporal evolution of magnetic imprints during GOES X-class flares. Top two events (a-d) are eruptive and bottom two (e-h) are non-eruptive events. Left column: time evolution of average $B_h$ (black) and the total Lorentz force, $F_z$ (blue) over the ROI along with their uncertainties. Right column: The corresponding time evolution of total magnetic energy (black), potential energy (green), and free-energy (blue). The pale yellow shaded region corresponds to the uncertainties in free-energy. Note the step-wise changes are co-temporal with flare timings. In all panels, vertical dashed lines refer to flare start (black), peak (red) and end times (black) respectively. 
}
\label{fig2}
\end{figure*}
%%%%%%%%%%%%%%%%%%%%%%%%%%%%%%%%%	

\citet{Fisher2012} provided the approximate expression for the total Lorentz force acting on the atmospheric volume enclosing the flaring AR. The equation for the vertical component of total Lorentz force ($F_{z}$) is approximated to be the surface integral of magnetic field over the photospheric surface (the upper and lateral boundaries are considered as sufficiently far to ignore their magnetic field contributions) and is given by, 
   \begin{equation}
	F_z = \frac{1}{8\pi} \int_A ( B_z^2 - B_h^2) dA
	\label{eq_fz}              
	\end{equation}
Where $dA$ is the pixel area ($1.3 \times 10^{15} cm^{2}$). \citet{Fisher2012} asserted that the total Lorentz force acting outward on the exterior from the interior has the same magnitude but opposite sign as the total Lorentz force acting inward on the interior from the exterior. Using the equation~\ref{eq_fz}, the Lorentz force impulse or changes in the Lorentz force vector acting on the volume at and below the photosphere during flare can be determined. If we have the photospheric vector magnetic field observations at two times, t=0 the start time of the flare and $t=\delta t$ the end time of the flare, then the changes in the downward Lorentz force ($\delta F_{z}$) during the flare is given by \citep{Hudson2008,Fisher2012}:
	\begin{equation}
	\delta F_z = \frac{1}{8\pi} \int_A ( \delta B_z^2 - \delta B_h^2)\,dA
	\label{eq_dfz}              
	\end{equation}
It is worth to note that $B_{z}$ shows no rapid irreversible change in any of our sample events during flares. Therefore, we omitted the term $\delta B_z^2$ in equation~\ref{eq_dfz} while performing computations. Neglecting $\delta B_z^2$ in equation~\ref{eq_dfz} means that the integrand is essentially the same as that in the integral of $\delta B_h$, except the quadratic scaling in the force integral. Consequently, all differences between the integral of $\delta B_h$ and $\delta F_z$ will only reflect this difference in scaling. The summation of $\delta F_z$ in the whole ROI gives the value of the integrated Lorentz force change and the computed values for sample events are tabulated in column 8 of Table~\ref{Tab1}. The temporal evolution of horizontal magnetic field and total downward Lorentz force in the ROI are discussed in the subsection~\ref{mag_imp}.
	
\subsection{Magnetic energy}
\label{me}
Given the vector magnetic field observations at the photospheric surface, the total magnetic energy of the AR can be estimated using the virial theorem \citep{Chandrasekhar1961, Molodensky1974, Low1982} and is given by                                      
	\begin{equation}
	E = \frac{1}{4\pi} \int_S (x B_x + y B_y) B_z dx dy  
	\label{eq_me}              
	\end{equation}
The assumption involved in the derivation of the above equation is that the photospheric magnetic field is force-free, which is not the case with the observations. Also, finite area surrounding the AR of interest is not fully flux-balanced. Though there has been no solid evidence that the virial energy estimate is proportional to magnetic energy in non-force-free fields (NLFFF), we speculate that virial energy estimate serves as a valid proxy for total magnetic energy in NLFFFs. %The virial energy estimates have unknown systematic errors, which are difficult to estimate. %So, the virial estimate of photospheric energy just acts as a proxy to the magnetic energy of ARs \citep{Su2014}. 
It's worth to note that the improved virial energy estimate method \citep{Wheatland2006} provides origin-independent energy estimates. We adopted the fourier transform method \citep{Gary1989} to derive the potential magnetic field using the $B_z$ component as input. The minimum energy state is attributed to current-free or potential field state and any energy excess to it is considered as free-energy or non-potential energy. Thus by subtracting the potential energy ($E_p$) from the total energy ($E$), gives the free energy ($E_f$) available in the AR for the energetic events likes flares and CMEs. The temporal variations of total, potential and free energy of our sample events are discussed in subsection~\ref{v_fe}.

\begin{figure*}[!ht]
	\centering
  \includegraphics[width=.98\textwidth,clip=]{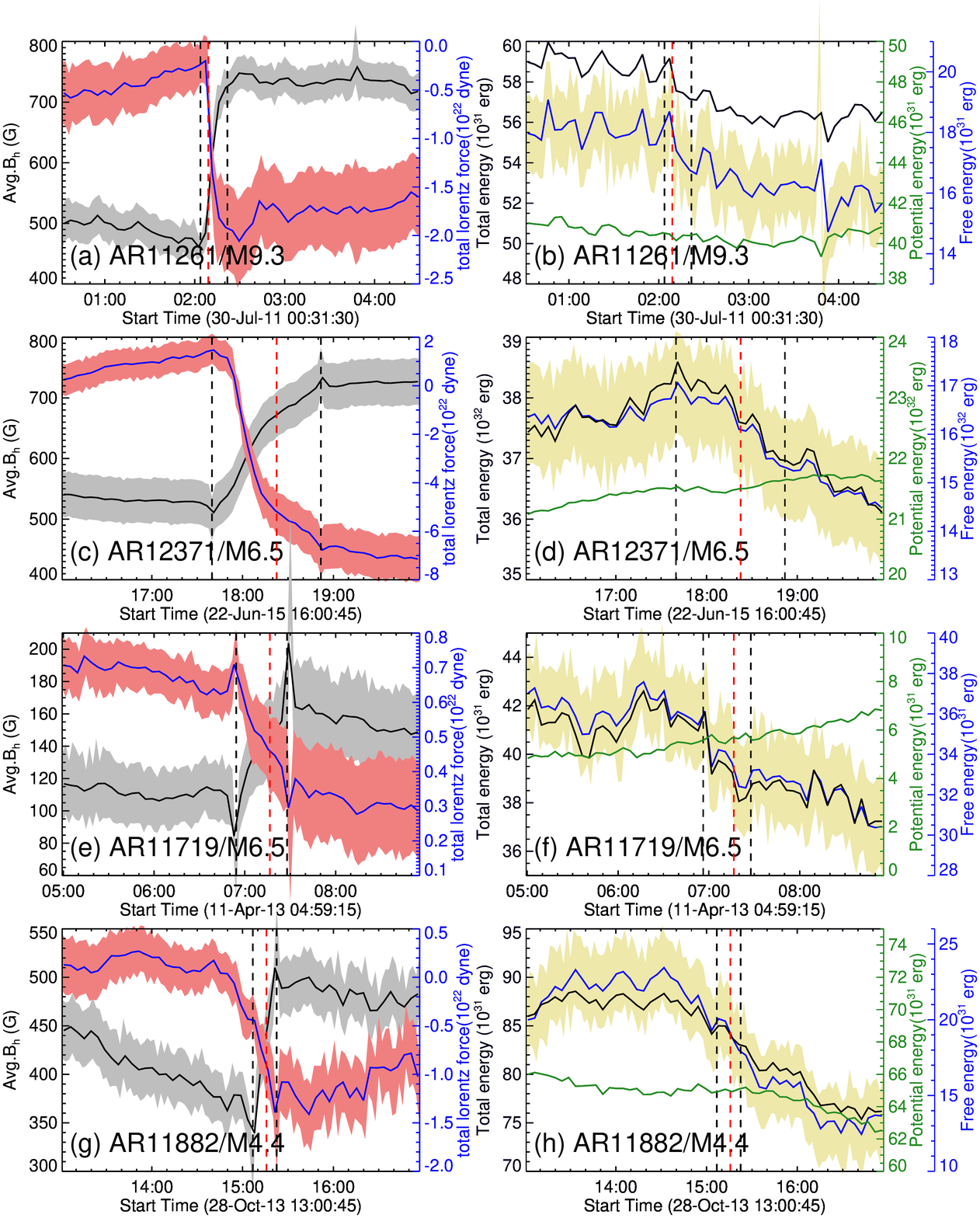}
	\caption{Same as in figure~\ref{fig2} but for GOES M-class flare events. M9.3 event from AR 11261 is non-eruptive event (a-b) and remaining events are eruptive (c-h).}
	\label{fig3}
\end{figure*}

\begin{figure*}[!ht]
	\centering
	\includegraphics[width=.98\textwidth,clip=]{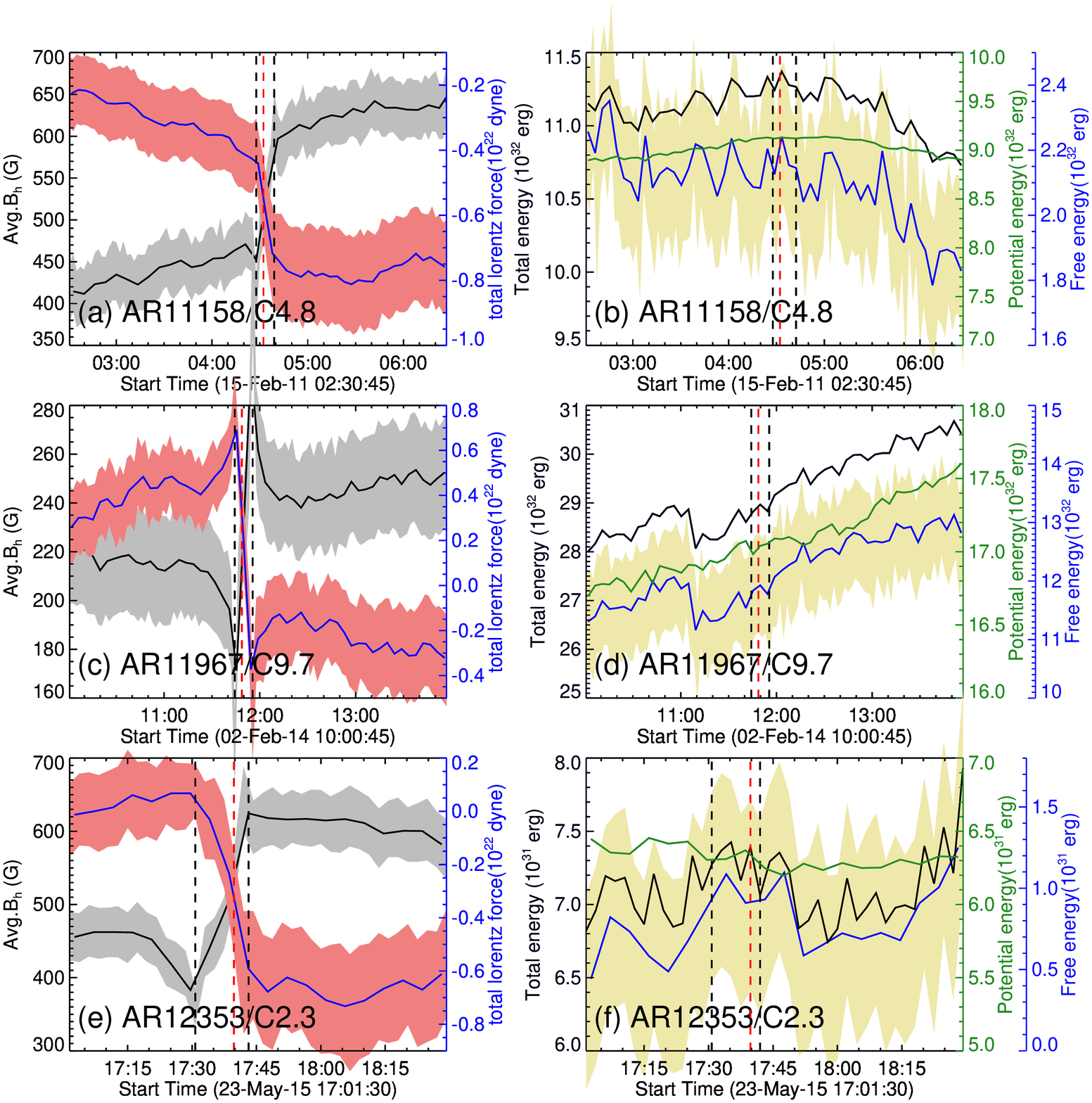}
	\caption{Same as in figure~\ref{fig2} but for GOES C-class flare events (all three are non-eruptive events).} 
	\label{fig4}
\end{figure*}

%\begin{figure*}[!ht]
%	\centering
%	\includegraphics[width=.98\textwidth,clip=]{fig5_ene_flux_ar11429}
%	%\includegraphics[width=.48\textwidth,clip=]{ene_ar12192}
%	\caption{The temporal evolution of free energy during an equivocal flare event of M6.3 from AR 11429 in our sample. The increase of free energy from the flare peak time seems to behave quite opposite to that of the events presented in figure~\ref{fig3}, whereas unvarying potential energy behaves same as the remaining events.}
%	\label{fig5}
%\end{figure*}

%\begin{figure*}[!ht]
%	\centering
%	\includegraphics[width=.98\textwidth,clip=]{fig6}
%	%	\includegraphics[width=.88\textwidth,clip=]{fig4b}
%	%	\includegraphics[width=.88\textwidth,clip=]{fig4c}
%	\caption{Temporal evolution of total unsigned electric current during eruptive and non-eruptive flares. Left column is for eruptive events and right column is for non-eruptive events.}
%	\label{fig6}
%\end{figure*}

%\begin{figure*}[!ht]
%	\centering
%	\includegraphics[width=.98\textwidth,clip=]{fig7}
%	%	\includegraphics[width=.88\textwidth,clip=]{fig4b}
%	%	\includegraphics[width=.88\textwidth,clip=]{fig4c}
%	\caption{Same as in figure~\ref{fig6} but for $\alpha_{av}$}
%	\label{fig7}
%\end{figure*}

\section{Analysis and Results}
\label{Res}
	
\subsection{Temporal evolution of $B_h$ and $F_z$}
\label{mag_imp}
\begin{figure*}[!ht]
	\centering
	\includegraphics[width=.95\textwidth,clip=]{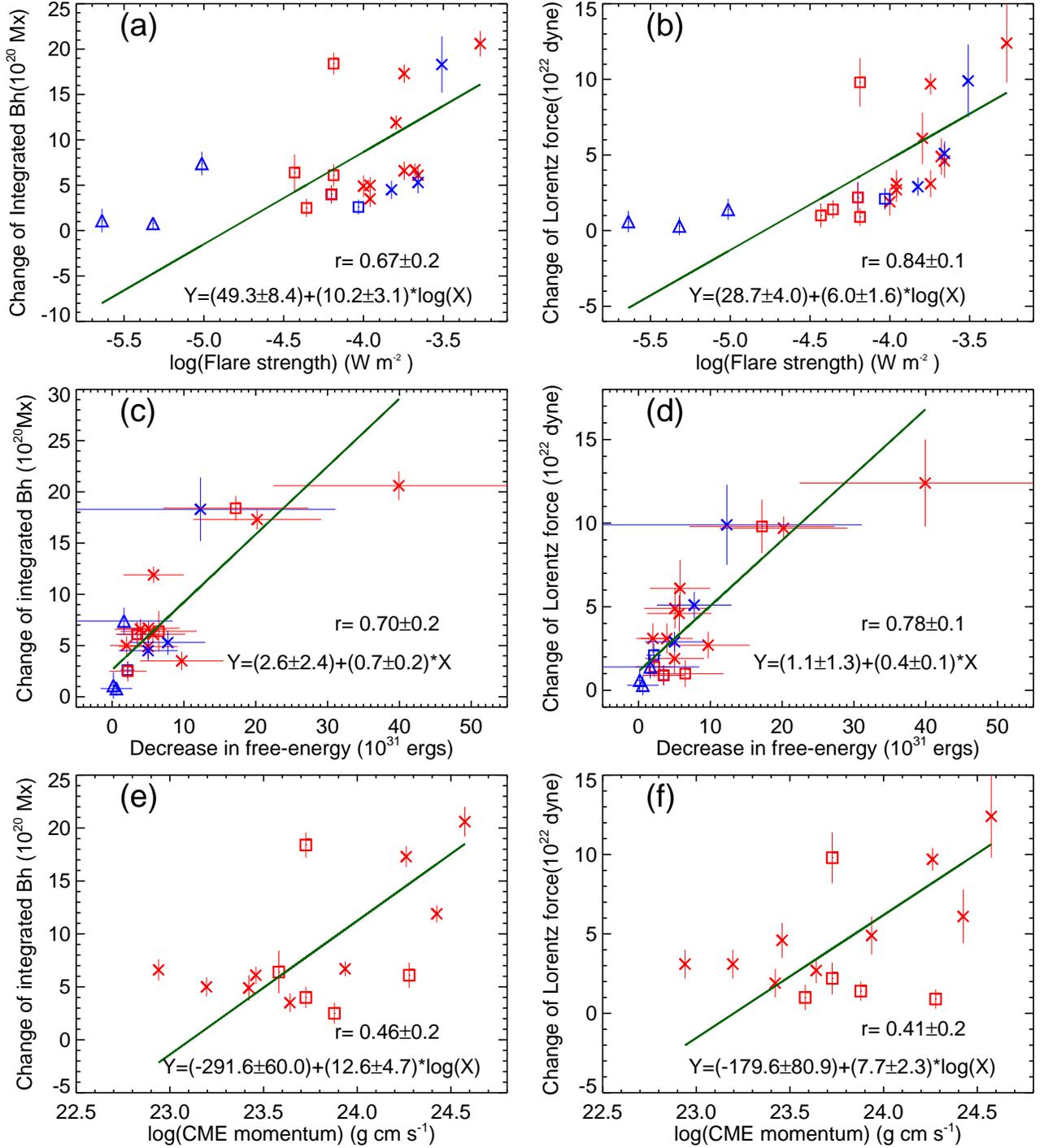}
	\caption{ Statistical relation between different physical parameters deduced from magnetic imprint regions.  
		Scatter plots of change of average $B_h$, and $\delta F_z$ against flare-strength (a-b), against free-energy decrease (c-d), against CME linear momentum (e-f). Note the X-axes in (a), (b), (e) and (f) are in logarithmic scale. The Spearman ranking correlation coefficient (r) and the resulting equation for straight line fit are given in the respective panels. The cross, square and triangle symbols represent the X-class, M-class and C-class flares respectively. Red (blue) colors of these symbols correspond to eruptive (confined) flare cases.}
	\label{fig5}
\end{figure*}

As described in the subsection~\ref{lf}, we carefully identified the ROI for each flare event in the sample. We studied the temporal evolution of average $B_h$ and total downward Lorentz force $F_z$ in the ROI by using high cadence 135 s vector magnetograms for a time window of 4 hours for each event (i.e., $\pm2$ hour from the flare peak time). The temporal evolution of average $B_h$ and $F_z$ for typical example events of four X-class flares (top two rows are eruptive and bottom two rows are non-eruptive) are plotted in the left-columns of Figure~\ref{fig2}. A similar evolution for M-class (top row is non-eruptive flare and remaining three rows are eruptive) and C-class (all three are non-eruptive) flares are plotted in Figure~\ref{fig3} and Figure~\ref{fig4} respectively. For strong field regions, the errors are minimum for HMI 135 s data  when compared to weak field regions \citep{Sun2017}. Hence, we assumed a linear variation ($y = -0.03 x + 74.44$) of errors (y) from 70 G to 30 G in the $B_x$(x) and $B_y$(x) values ranging from 150 to 1500 G, respectively. For the  $B_x$ and $B_y$ values having greater than 1500 G, a uniform error of 30 G has been assumed in $B_x$ and $B_y$. We then propagated these field component errors into the respective equations to finally obtain the errors in the magnetic imprint quantities. We estimated errors on all our computed quantities and plotted in Figures~\ref{fig2}, ~\ref{fig3}, ~\ref{fig4} and ~\ref{fig5}.

Both $B_h$ and $F_z$ are observed to vary in step-wise manner during all flare events. In these plots, three vertical dashed lines refer to flare start (black), peak (red) and end times (black) respectively. The horizontal magnetic field rapidly becoming more horizontal during the flare interval (inclination angle rapidly decreases; plot not shown) and this enhancement of $B_h$ is permanent which is in agreement with the past studies \citep{Wang2012b,Petrie2012,Sun2017,Zekun2019}. It is worth to point here that about 24 \% (5 events) of flare events in our sample are short-time duration events of $\leq$ 12 m. The average $B_h$ enhancement from their pre-flare values for short duration events ranges from $32\%$ to $52\%$. As an illustration (Figure~\ref{fig2}c), the flare X1.1 from NOAA 11890 have the time duration of 10 minutes and during this short interval, we observed the average $B_h$ enhancement of above 250 G ($>34\%$) from its pre-flare value. Such rapid changes were able to be recorded more precisely with the help of high cadence 135 s vector magnetograms than the routine version of 720 s cadence vector magnetograms of HMI. The remaining 76 \% of flare events in our sample are long duration events ranges from 13 m to more than a hour. These events exhibits the average $B_h$ enhancement from their pre-flare values in the range of $24\% - 55\%$ . This shows that the amount of $B_h$ enhancement from its pre-flare value does not depends much on the duration of flares. However, for longer-duration flares, true flare associated $B_h$ changes will likely be contaminated by noise from non-flare related evolution. Unfortunately, we couldn't quantify the noise estimation from non-flare related evolution for all the longer duration flares because HMI 135 s data has been processed for selective time intervals of prominent activity\footnote{\url{http://jsoc.stanford.edu/data/hmi/highcad/}}, leads to unavailability of non-flaring interval data. Nonetheless, the noise estimated for couple of longer duration flares in our sample indicates that the contribution of noise effect from non-flare related evolution is likely small than the true flare related $B_h$ changes. It's worth to note a caveat along with the latter is that the strength of noise from non-flare related evolution is very subjective.

%In general, our sample events exhibit a $30\%$ to above $50\%$ increase of the average $B_h$ from their pre-flare values during the flaring period. %\pv{better tell about long duration events, since you mention about short ones}. 

On the other hand, the total downward Lorentz force shows the abrupt decrease during the flare interval and this change is also irreversible. Here the negative sign assigned to the total Lorentz force represents only the direction i.e., the force imposed on the photosphere from above as indicated in equations~\ref{eq_fz} \& \ref{eq_dfz}. The absolute value of total change of Lorentz force gives the reasonable indicator of strength of magnetic imprints. The computed total $\delta F_z$ for all the events are tabulated in column 9 of Table~\ref{Tab1}. As indicated by the strength of magnetic imprints tabulated in columns 8 and 9 of Table~\ref{Tab1}, it is found that non-eruptive flares have comparable strength of magnetic imprints with that of eruptive counterparts. These results suggest that strength of magnetic imprints do not depend on whether the flare is eruptive or not.
% However, it is hard to arrive at the solid conclusion that whether or not the strength of magnetic imprints for eruptive flares is stronger than the non-eruptive ones from statistically small sample events.

\subsection{Variation of Magnetic Free energy}
\label{v_fe}
As described in section~\ref{me}, we computed the total magnetic energy and potential energy using the virial theorem and the proxy for magnetic free energy is obtained by their difference. For error estimation, we followed the same procedure as described in previous sub section~\ref{mag_imp}, the errors in the components of magnetic field are propagated through the Virial theorem to obtain the uncertainties in energy estimates. For greater clarity, only uncertainties in free energy are showed in Figures~\ref{fig2}, \ref{fig3} and \ref{fig4} as pale yellow shaded regions. We used the 135 s cadence vector magnetogram data for the time period of 4 hours ($\pm2$ hour from the flare peak time). For each event, we choose the entire AR such that the flux balance within the region is as good as possible to compute the potential field and to deduce the magnetic energy.
%, we choose the region such that it encloses only the eruption region of AR and flux balance within the region is as good as possible \pv{this arises controversies },
Decrease in free energy during several flares in our sample are not statistically significant, but in the 14 out of 21 events that exhibit decrease in free-energy exceeds 1 sigma error level. The temporal evolution of all three magnetic energies for typical events of X-class, M-class and C-class are plotted in right column panels of Figure~\ref{fig2}, ~\ref{fig3} and~\ref{fig4} respectively. As the potential field only depends upon $B_z$ component, which is basically unchanged during the given period of evolution, potential energy represented by green curve in all these panels also remains constant. Consequently the free energy in blue and total energy in black exhibit the same decreasing trend during and after the flares. These results are in agreement with the past studies of individual events \citep{Ravindra2010, Vemareddy2012} and statistical studies (for eg. \citealp{Vasantharaju2018}). The difference in free energy at the flare start and end times as indicated by the two vertical black dashed lines in all panels of Figure~\ref{fig2},~\ref{fig3} and~\ref{fig4} gives the amount of decrease in free energy ($\Delta FE$). This $\Delta FE$ accounts for energy release during flares. The computed values of $\Delta FE$ for these 20 events are tabulated in column 9 of table~\ref{Tab1}. Estimates of errors for the change of quantities : integral of $\delta B_h$, $\delta F_z$ and $\Delta FE$ are also provided in Table~\ref{Tab1}. The largest amount of free-energy released in our sample is about $4\times10^{32}\,ergs$ during the X5.4 flare event and giving out fastest CME in our sample with a linear speed of 2684 km s$^{-1}$. The range of amount of free energy released during X-class flare events in our sample is from $2\times10^{31}$ to $4\times10^{32}$ ergs. Similarly, for M-class flares : $2\times10^{31}$ to $1.7\times10^{32}$ ergs and for C-class flares : $1.8\times10^{30}$ to $1.6\times10^{31}$ ergs. We note that the amount of free energy released during non-eruptive flares are comparable to that of eruptive counterparts. Thus, we opine that the amount of free-energy released during a flare have minimum dependency on whether the flare being eruptive or not. The fluctuations in the free energy seem to be dominated by fluctuations in the total energy. % and in a statistical sense, the decrease in free energy for majority of our events are not significantly larger than the fluctuations.
%\pv{explain ranges of free-energy decrease for different flare events, }
	
As an exceptional case in our sample, we did not observe the decrease in free energy during M6.3 flare from AR 11429. On a contrary, we did find the free energy starts increasing from the flare peak time. 
%The temporal evolution of magnetic energies in AR 11429 during M6.3 flare is shown in Figure~\ref{fig5} (a). 
AR 11429 is a highly complex $\beta \gamma \delta$ region produced a sequence of three recurrent M-class flares during 3:00 - 4:00 UT on 9 March 2012 \citep{Simoes2013, Polito2017}. The eruptive M-class white light flare started at 03:22 UT, reached a first maximum in the soft X-ray flux at around 03:27 UT (M1.8 class) and after a small dip in intensity, it rose to M2.1 class at 03:37 UT and further with a small dip, it increased up to about M6.3 class at 3:53 UT. Being long duration and recurrent flare event, there is a probability of increase in magnetic free energy due to the dynamic activity of AR and this is evident in the increase of total flux of AR during the flare (plot not shown). %\pv{not understand this statement} 
Mostly, this masks the decrease in free energy during the flare. Thus this interesting event is a matter of subjective study and a comprehensive analysis of this event is beyond the scope of our motivations in the present study.

\subsection{Statistical results of magnetic imprints}
\label{stats}
To find the relationship between the magnetic imprints and the eruptive properties, we performed the linear regression analysis on the scattered plots shown in Figure~\ref{fig5}). \texttt{FITEXY} routine from SolarSoft library has been employed to carry out the linear regression analysis. This routine is based on least-square approximation to fit the best straight line and the standard deviation of both x $\&$ y data coordinates are used as error inputs. The slopes and y-intercepts along with their uncertainties obtained from the regression model are shown in the equation of respective panels of Figure~\ref{fig5}. The Spearman ranking correlation coefficient ($r$, referred as CC) is estimated in all our studies and the  standard error in $r$ is estimated by  $ERR_r = \sqrt ((1-r^2)/(n-2))$. 

We plotted the log of flare strength against the integrated change of $B_h$ and total change of $F_z$ in Figure~\ref{fig5}(a) $\&$~\ref{fig5}(b) respectively. The cross, square and triangle symbols represent the X-class, M-class and C-class flares respectively. From these scatter plots, good correlation is realized (flare strength has CC of 0.67 with integrated change of $B_h$ and CC of 0.84 with change of $F_z$) and suggest that stronger flares tend to produce more horizontal field changes and these changes result in stronger downward impulse on the photosphere. This result is in agreement with \citet{Wang2010} and \citet{Wang2012b}. However, it is worth to note here that two of our eruptive flare events of same GOES class strength of M6.5 occurred from NOAA 11719 and NOAA 12371 produce varying strength of magnetic imprints on photospheric magnetic field. The M6.5 from NOAA 12371 produces the stronger strength of magnetic imprints than the M6.5 from NOAA 11719 (see Table~\ref{Tab1}). Thus, strength of the flare may not be the only indicator of the magnitude of the magnetic imprints produced. Also, we do not see any evident relationship between the strength of magnetic imprints and the eruptive/non-eruptive nature of flares. The blue (red) color symbols correspond to  non-eruptive (eruptive) flares in  Figure~\ref{fig5}. It is noted that the strength of magnetic imprints produced is less likely dependent on the eruptive or non-eruptive nature of flares but more likely depends on the strength of the flares.

The Figure~\ref{fig5}(c) and~\ref{fig5}(d) display the scatter plots for the amount of the free energy decrease against the integrated $\delta B_h$ and total $\delta F_z$ respectively. We excluded the event M6.3 from AR 11429 in the scatter plot as it exhibits the increase in free energy during flare. The $\Delta FE$ has CC of 0.70 with integrated $\delta B_h$ and 0.78 with total $\delta F_z$ referring to a good correlation. This indicates that the amount of energy released in the corona during flares is strongly related to the photospheric field changes and in turn the downward Lorentz force impulse on the photosphere. It means that larger the amount of free energy released during flares, more the strength of magnetic imprints produced. From the scatter plots, it is also evident that stronger flares release larger amount of energy irrespective of being eruptive or not and in turn produces stronger strength of magnetic imprints.

\begin{figure*}[!ht]
	\centering
	\includegraphics[width=.99\textwidth,clip=]{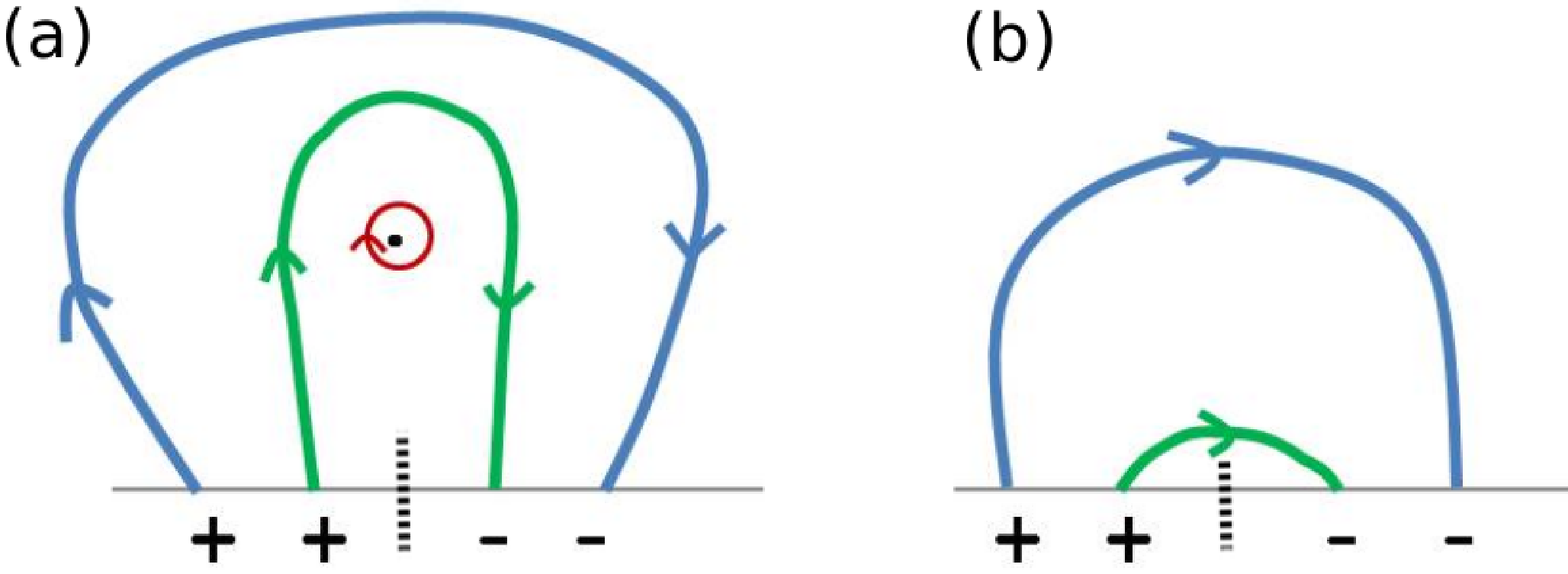}
	\caption {Schematic representation of core magnetic field becoming ``more horizontal" after the implosion in a flaring AR. (a) During pre-eruption, the core field enclosing the flux rope (red circle) indicated by green curve is mostly vertical in nature. While the peripheral magnetic field lines, represented by blue curve are tilted away from the core or roughly horizontal in nature. Whereas in post-eruption scenario as depicted in (b), implosion makes the core field to become more horizontal and peripheral field to be more vertical.}
	\label{fig6}
\end{figure*}

%It means that part of the energy released during flares/CMEs reaches solar interior via back reaction. This result favors the idea of coupling of flare energy into a seismic wave identified as magnetic jerk in the solar interior, which is produced by the downward impulse in the vertical direction due to change of lorentz force during flares \citep{Anwar1993, Hudson2008}. 

Further to find the relationship between CME eruption and photospheric field changes during flares, we used the properties of 14 eruptive events in our sample. Based on the conservation of linear vertical momentum, \citet{Fisher2012} suggests that upward impulse from the Lorentz force change drives the magnetic eruption and the impulse can be related to CME linear momentum. CME linear momentum is computed using the CME mass (M) and linear speed (v) obtained from the Large Angle and Spectrometric Coronagraph (LASCO) CME catalog\footnote{\url{https://cdaw.gsfc.nasa.gov/CME_list/}}. The generated scatter plots are shown in the panels (e) and (f) of Figure~\ref{fig5}. The CME linear momentum has the CC of 0.46 with the integrated $\delta B_h$ and 0.41 with $\delta F_z$. The CME linear speed has the CC of 0.52 with the integrated $\delta B_h$ and 0.45 with $\delta F_z$ (plot not shown). It indicates that large change of Lorentz force leads to stronger upward impulse which drives stronger CMEs. This result is in agreement with the quite recent study of \citet{Zekun2019}. We attributed this weak correlation is mostly due to the poor CME mass estimates in CDAW CME catalog \citep{Vourlidas2002}. The involved assumptions in the CME-momentum estimate are that the entire mass of the CME structure moves with uniform speed and work done against gravity is neglected, which may differ from the ideal case of momentum change proportional to the product of average Lorentz force and the time period over which it acts.

\begin{figure*}[!ht]
\centering
		\includegraphics[width=14.7cm,height=5.2cm,clip=]{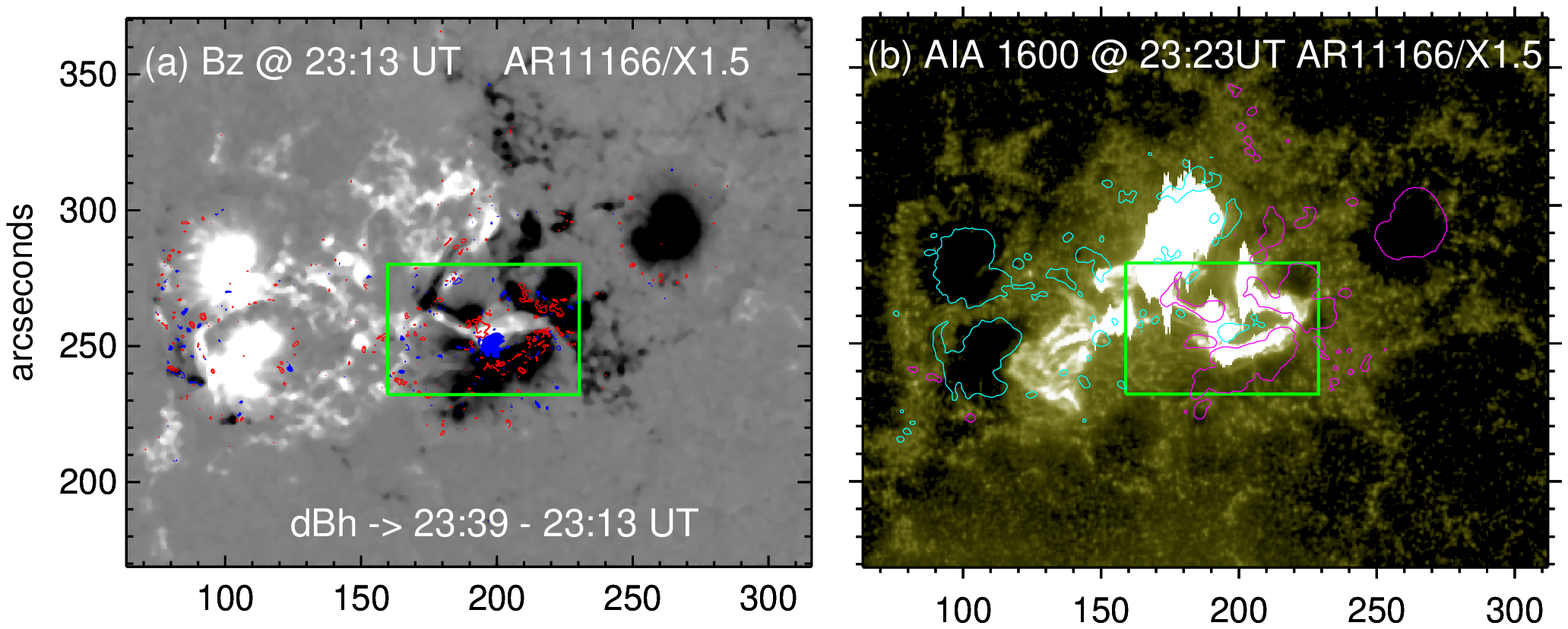}
		\includegraphics[width=15.9cm,height=10.2cm,clip=]{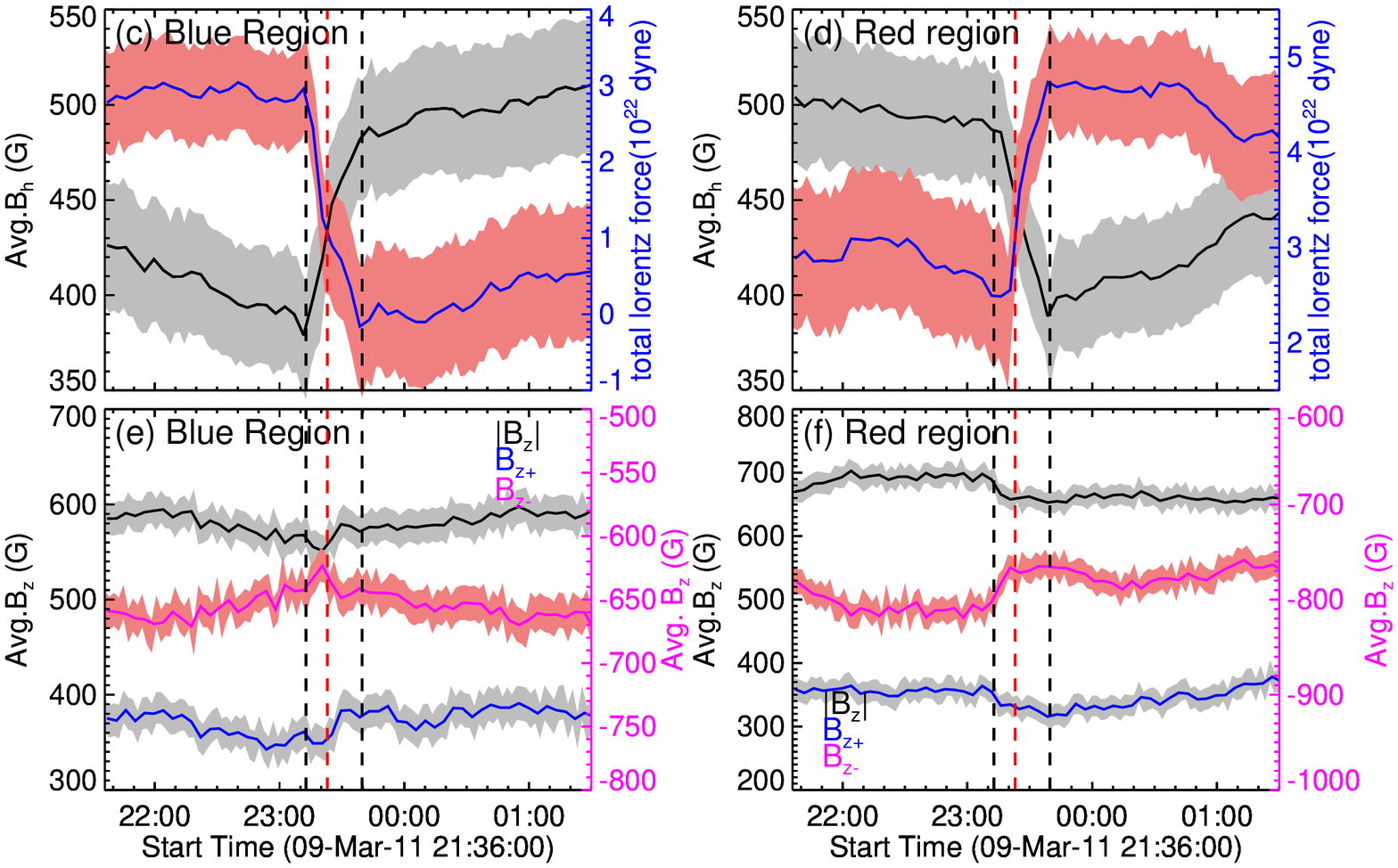}
\caption{An example of {\bf non-eruptive} flare event (X1.5 from AR 11166) to illustrate the cancellation of Lorentz forces. {\bf a)} $B_z$ map overlaid with the blue and red color contours of $B_h$ enhancements (200 G \& 300 G)  and decrements (-200 G and -300 G) during the flare start time (23:13 UT) and end time (23:39 UT) respectively. {\bf b)} AIA 1600~\AA~image observed at flare peak time (23:23 UT), overlaid are the $B_z$ contours at $\pm 600~G$. The green box in these two panels is the ROI where the flare kernels in panel (b) are spatially correlated with the enhancements and decrements of $B_h$ in panel (a). {\bf c)} temporal evolution of average $B_h$ and derived Lorentz force over the $B_h$ enhancement (blue) regions within the green box of (a). {\bf d)} Same as (c) but over the $B_h$ decremented (red) regions enclosed by the green box of (a), {\bf e-f)} temporal evolution of absolute $B_z$ and net flux in the blue and red regions within the green box of (a) respectively. Note the Lorentz forces are acting in opposite directions within the green box and variations of $B_z$ show no  predominant irreversible changes during the flare.}
\label{fig7}
\end{figure*}

\section{Summary and Discussion}  
\label{summ}
We studied the flare associated changes in photospheric magnetic field during the 21 flare events of varying GOES class strengths from C-class to X-class (14 eruptive and 7 non-eruptive events) from 17 ARs using quite recently released and selectively processed 135 s cadence HMI full disk vector magnetograms. The summary of main data analysis results obtained are as follows:
\begin{itemize}
\item[1.] We observed the rapid and permanent changes on the surface magnetic field in and around the flaring PIL regions of AR. The observed step-wise enhancement in horizontal field component leads to step-wise changes of Lorentz force over the manually selected ROIs. The average $B_h$ enhancement from their pre-flare values are in the range of $24\% - 55\%$. The MI regions produced by the eruptive flare events in our sample are strongly localised, whereas MI regions produced by the majority of non-eruptive events ($>70~\%$) are observed to be scattered.
\item[2.] The changes of horizontal field integrated over ROI and the magnitude of the downward impulse (produced from the change of Lorentz force), both termed as strength of magnetic imprints, are well correlated with the strength of flares having CC of 0.67 and 0.84 respectively. It is found that irrespective of flares being eruptive or not, short ($\leq 12\,minutes$) or long duration ($> 13\,minutes$), stronger flares tend to produce stronger strength of magnetic imprints on the surface magnetic field.  
\item[3.] The magnetic free energy released during the flares under study, estimated using the virial theorem over the whole ARs, show the strong correlation (CC = 0.78) with the downward impulse imposed on the photosphere. It is noted that stronger flares irrespective of whether they are eruptive or not, releases larger amount of free energy which in turn produces stronger strength of magnetic imprints.
\item[4.]  The magnetic free energy exhibits statistically significant downward trend which starts around the flare time is observed in majority (14 of 21) of our strong flares.
%Many strong flares (14 of 21) exhibited statistically significant downward trend in free energy that started around their flare times.}
\item[5.] Due to the lack of reliable CME mass estimates, we were constrained to find the relationship between CME linear speed and the strength of magnetic imprints produced. The positive correlation (CC $\approx 0.5$) between them indicates that stronger upward impulse, produced by larger change of Lorentz force, drives the CME faster. 
\end{itemize}

%The present study shows that the amount of free energy released are strongly related to the strength of magnetic imprints produced irrespective of eruptive or non-eruptive flares. 
In the same line of argument as that of result 3, the recent statistical study of \citet{Vasantharaju2018} showed that the released magnetic free energy is well correlated to the intensity of flares (including 38 eruptive and 39 non-eruptive flares). These studies imply that it is hard to use a criteria to distinguish between eruptive and non-eruptive flare events in terms of released magnetic free energy. 

Further, \citet{Hudson2008} and \citet{Fisher2012} suggested that pre-flare magnetic field configuration is under equilibrium of balanced forces. This equilibrium of force balance gets disrupted by the rapid release of coronal free energy during flares, the unbalanced Lorentz force change drives the eruption impulse in the upward direction and due to conservation of momentum there should be an equal and opposite impulse acting on the plasma of solar interior to produce the magnetic jerks or ``McClymont jerk". Also, \citet{Emslie1982} using their analytical model showed that considerable wave energy can be penetrated into the photosphere and disturbs the magnetic field there. \citet{Anwar1993} made a rough estimation of how much transmitted energy goes into seismic waves. In this study, we found the strong association of flare energy released with magnitude of Lorentz force change. This result favors the idea of coupling of flare energy into a seismic wave identified as magnetic jerk in the solar interior, which is produced by the downward impulse due to change of Lorentz force during the flares.
%CME momentum is well correlated with the magnitude of upward/downward impulse. 
These observational results and their relationship with the eruption properties reveal that the photospheric magnetic fields respond to the back reaction of coronal field restructuring due to flare energy release, which strongly agrees with the coronal implosion notion of \citet{Hudson2000} and \citet{Hudson2008}.

According to Coronal implosion model \citep{Hudson2000}, the core of the AR as indicated by green curve in Figure~\ref{fig6}, around the flaring PIL, contains a non-potential structure that exerts outward magnetic pressure on surrounding regions of the AR. This pressure can cause the surrounding fields (blue curve in Figure~\ref{fig6}(a)) to tilt away from the core. During a flare (eruptive or not), the core field loses energy and in turn  magnetic pressure decreases, and eventually the field becomes ''more compact'' as shown in green curve of Figure~\ref{fig6}(b). Whereas, the fields that were tilted away from the core before the flare could become "less tilted away" from the core or more vertical as depicted by blue curve in Figure~\ref{fig6}(b). However, many recent observational studies \citep{Kleint2017,Sun2017} questioned on the validity of the implosion. The model claims that contracting coronal structure leads to the increase of horizontal magnetic field in photosphere. But, \citet{Kleint2017} showed that chromospheric field do not evolve in coherent with the photospheric field. This indicates coronal field evolution need not be similar to photospheric field evolution and expresses doubt on the origin of magnetic imprints suggested by the model.

%Though these relationships are seem to be in strong agreement with the coronal implosion conjecture \citep{Hudson2000}, the observations of simultaneous increase and decrease of horizontal magnetic field in and around the flaring region cancelling oppositely directed Lorentz forces strongly contradicts the conjecture. Also, the conjecture explains that contracting coronal structure leads to the increase of horizontal magnetic field on photosphere. But, recent study \citet{Kleint2017} showed that chromospheric field do not evolve in coherent with the photospheric field. This indicates coronal field evolution need not be similar to photospheric field and  expresses doubt on the origin of magnetic imprints suggested by the conjecture. Recently \citet{Sun2017} suggested that upward impulsive Lorentz force should not be confused with the downward force exerted on and below the photosphere causing the ``sunquakes''. 

Also, \citet{Fisher2012}'s model which explains the relationship between magnetic eruption and the photospheric magnetic field changes over the course of a solar flare, contradicts our observations. Following is the illustration of such a contradictory observation with one of our sample events. During the X1.5 flare from AR NOAA 11166, we observed the increase and decrease of $B_h$ values of more than 150 G (upto 350 G) during the flare over almost entire AR and are represented by blue and red color contour regions respectively, as shown in fig~\ref{fig7}a. The increment and decrement regions of $B_h$ are usually observed to occur adjacent to each other. To study the flare associated changes in AR, the green box in the panels (a) \& (b) of Figure~\ref{fig7} are chosen in such a way that flare kernels should be spatially coherent with the increments and decrements of $B_h$. The $B_h$ increment regions are prominently observed near the flaring PIL and the $B_h$ decrement regions are mostly occurred around the polarity regions. The temporal evolution of $B_h$ and the derived Lorentz force in these regions are shown in panels (c) \& (d) of Figure~\ref{fig7}. These plots imply that the co-temporal increase and decrease of $B_h$ results in the Lorentz forces acting in opposite directions simultaneously. The total positive change of Lorentz force ($3.24\times10^{22}$ dyne) due to the increase of $B_h$ gets canceled by the negative change of Lorentz force ($-3.09\times10^{22}$ dyne) due to decrease of $B_h$ within the volume lying above the surface region enclosed by the green box in Figure~\ref{fig7}. Whereas, the $B_z$ variations during the flare as shown in panels (e) \& (f) of Figure~\ref{fig7} didn't exhibit any flare associated permanent changes. Thus, the increase and decrease of $B_h$ leads to the cancellation of oppositely directed Lorentz forces and the net upward impulse will be insufficient to drive the CME momentum, makes a contradiction to the model proposed by \citet{Fisher2012}. 

Likewise, many observations related to magnetic imprints yet to understand completely. We observed the increase (step-wise) of non-potential parameters like total current, twist parameter ($\alpha$), magnetic shear/weighted shear angle and the horizontal component of magnetic field during flares. As the vertical magnetic field remains constant through out the flare, we expect the magnetic free energy to increase (according to equation~\ref{eq_me}) but actually it decreases during flares. Though there were attempts explaining how the magnetic field becomes more horizontal during flares using tether cutting reconnection model (eg.\citealp{Wang2012a}) or enhancement of non-potentiality due to emergence of sheared fields \citep{Jing2008}, by and large the origin of magnetic imprints is still an open question. These contradictions indicate the need of a general model which helps us to better understand the origin of magnetic imprints.

\acknowledgements Authors thank the reviewer for insightful comments and suggestions which greatly improves the quality of the manuscript. The version of cartoon concept which is employed as Figure 6 in the manuscript was initially suggested by the reviewer. Authors thank Dr. Manolis K. Georgoulis, AAS Journals Scientific Editor for comments and suggestions which further improves the presentation of the manuscript. SDO is a mission of NASA's Living With a Star Program. We thank the HMI science team for the open data policy of high cadence vector magnetograms. The CME catalog is generated and maintained at the CDAW Data Center by NASA and The Catholic University of America in cooperation with the Naval Research Laboratory. SOHO is a project of international cooperation between ESA and NASA.
\bibliographystyle{apj}
\bibliography{mi_ref}

\end{document}